\def\be{\begin{equation}}
\def\ee{\end{equation}}
\def\ba{\begin{array}}
\def\ea{\end{array}}

\def\Cb{{\Bbb C}}
\def\cb{{\Bbb C}}

\documentstyle[11pt]{article}
\input amssym.def
\def\ra{\rangle}
\def\la{\langle}

\begin{document}
\baselineskip18pt
\thispagestyle{empty}
\begin{center}
{\LARGE \bf  Separability and entanglement in $\cb^2 \otimes
\cb^3 \otimes \cb^N$ composite quantum systems}
\end{center}
\vskip 2mm

\begin{center}
{\normalsize Shao-Ming Fei$^{1,\  2}$, Xiu-Hong Gao$^1$, Xiao-Hong
Wang$^1$, Zhi-Xi Wang$^1$ and Ke Wu$^1$ }
\end{center}

\begin{center}

\begin{minipage}{4.8in}
{\small \sl $^1$
Department of Mathematics, Capital  Normal University, Beijing 100037,
China}

{\small \sl $^2$ Institute of Applied Mathematics,
University of Bonn,  53115 Bonn, Germany}
\end{minipage}
\end{center}
\vskip 3mm

\begin{center}
\begin{minipage}{5in}
{\bf Abstract}
The separability and
entanglement of quantum mixed states in  $\Cb^2 \otimes \Cb^3
\otimes \Cb^N$ composite quantum systems are investigated.
It is shown that all quantum states $\rho$ with positive partial transposes
and rank $r(\rho)\leq N$ are separable.
\end{minipage}
\end{center}
\vskip 4mm

\section{Introduction}

As one of the most striking features of
quantum phenomena \cite{1p}, quantum entanglement is
playing very important roles in quantum information processing
such as quantum computation \cite{DiVincenzo}, quantum
teleportation \cite{teleport,teleport1,Ari00,teleport2} (for
experimental realization see \cite{telexp}), dense coding
\cite{dense} and quantum cryptographic schemes
\cite{crypto1,crypto2,crypto3}. The separability of pure
states for bipartite systems is quite
well understood (cf. \cite{peresbook}). Nevertheless,
The study of separability and entanglement of quantum mixed states
is far from being satisfied. A mixed state
is considered to be entangled if it is not a mixture of product
states \cite{10}. As the quantum correlations are
weakened in mixed states, the manifestations of mixed-state entanglement can
be very subtle \cite{10, 11, 12}. To investigate the structure of
mixed state entanglement some beautiful works have been done in
quantifying entanglement \cite{17, 18, 19, 20, 21, fjlw} for bipartite
systems and multipartite systems (see e.g. \cite{22,23}).
However most proposed measures of
entanglement for bipartite systems involve extremizations which
are difficult to handle analytically.
For multipartite systems, one even does not know how to define
the measures. Till now there is no general criterion that allows one to
distinguish whether a mixed state is separable or not.

Some progress has been achieved in understanding the separability and
entanglement problem for bipartite systems (cf. \cite{primer}),
e.g., the proper definition of separable and entangled states
formulated by Werner \cite{10}, the Peres \cite{Peres}
criterion that all separable states necessarily have a positive
partial transpose (PPT), which is further shown to be also a sufficient
condition for separability in $2\times2$ and $2\times 3$ systems
\cite{ho96,tran}. There have been many results on the separability
and entanglements of mixed states, see e.g., \cite{book,1,2,3,we}.
Recently some new criterion are presented
in \cite{chenkai,rudolph}, which are necessary conditions for
a state to be separable and complement the
well-known PPT criterion in certain aspects.

In \cite{2}, the separability and
entanglement of quantum mixed states in $\Cb^2 \otimes \Cb^2
\otimes \Cb^N$ composite systems are investigated and
some very interesting results were obtained.
In this paper we generalize these results to
quantum mixed states in $\Cb^2 \otimes \Cb^3
\otimes \Cb^N$ composite systems. It is shown that
all quantum states $\rho$ with positive partial transposes
and rank $r(\rho)\leq N$ are separable.
In the following we denote by $R(\rho)$, $K(\rho)$, $r(\rho)$ and
$k(\rho)$ the range, kernel, rank and the dimension of the
kernel of $\rho$, respectively. The three subspaces
of $\Cb^2 \otimes \Cb^3\otimes \Cb^N$ will be called
Alice, Bob and Charlie.

\section{The generic form of rank-$N$ PPT states}

We first consider the canonical form of a separable state in
$\Cb^2 \otimes \Cb^3 \otimes \Cb^N$ with $r(\rho)=N$.

{\bf Lemma 1. }\  Every PPT state $\rho$ in $\Cb^2 \otimes \Cb^3
\otimes \Cb^N$ with $r(\rho)=N$ such that in some local basis
$\{|0_A\ra, |1_A\ra\}$ for Alice,, $\{|0_B\ra, |1_B\ra, |2_B\ra\}$
for Bob, $\{|0_C\ra \cdots |N-1_C\ra\}$ for Charlie without losing
the generality we assume $r(\la 1_A, 2_B|\rho |1_A, 2_B\ra)=N$,
can be transformed, by using a reversible local operation, into
the following canonical form: \be\label{lemma1} \rho=\sqrt{F}
\left(
\begin{array}{c}
C^{\dag}D^{\dag}\\
B^{\dag}D^{\dag}\\
D^{\dag}\\
C^{\dag}\\
B^{\dag}\\
I
\end{array}
\right)
\left(
\begin{array}{cccccc}
DC&DB&D&C&B&I
\end{array}
\right)\sqrt{F},
\ee
where $[B,\ B^{\dag}]=[C,\ C^{\dag}]=[D,\ D^{\dag}]=[C,\ B]=[C,\
B^{\dag}] =[C,\ D]=[C,\ D^{\dag}]=[B,\ D]=[B,\ D^{\dag}]=0$ and
$F=F^{\dag}$; $B$, $C$, $D$, $F$ and the identity $I$ are
operators acting in the Charlie's space.

{\bf Proof. }\ In the considered basis the state $\rho$
can be always written as:
$$
\rho=\left(
\begin{array}{cccccc}
E_1&E_{12}&E_{13}&E_{14}&E_{15}&E_{16}\\
E_{12}^{\dag}&E_2&E_{23}&E_{24}&E_{25}&E_{26}\\
E_{13}^{\dag}&E_{23}^{\dag}&E_{3}&E_{34}&E_{35}&E_{36}\\
E_{14}^{\dag}&E_{24}^{\dag}&E_{34}^{\dag}&E_4&E_{45}&E_{46}\\
E_{15}^{\dag}&E_{25}^{\dag}&E_{35}^{\dag}&E_{45}^{\dag}&E_5&E_{56}\\
E_{16}^{\dag}&E_{26}^{\dag}&E_{36}^{\dag}&E_{46}^{\dag}&E_{56}^{\dag}&E_6
\end{array}
\right),
$$
where $E's$ are $N \times N$-matrices and $r(E_6)=N$. The
reduced matrix $\tilde{\rho}=\la 1_A|\rho|1_A\ra$ is of the form,
\[
\tilde{\rho}=\left(
\begin{array}{ccc}
E_4&E_{45}&E_{46}\\
E_{45}^{\dag}&E_5&E_{56}\\
E_{46}^{\dag}&E_{56}^{\dag}&E_6
\end{array}
\right).
\]
$\tilde{\rho}$ is a PPT state in $\Cb^3 \otimes \Cb^N$ and $r(\tilde{\rho})=r(\rho)=N$. 
Using Lemma 4 in \cite{1} we have
\[
\tilde{\rho}=\left(
\begin{array}{ccc}
C^{\dag}C&C^{\dag}B&C^{\dag}\\
B^{\dag}C&B^{\dag}B&B^{\dag}\\
C&B&I
\end{array}
\right),
\]
where $[B,B^{\dag}]=[C,C^{\dag}]=[C,B]=[C,B^{\dag}]=0$.

Similarly, the projection $\bar{\rho}=\la 2_B|\rho|2_B\ra$
gives rise to
\[
\bar{\rho}=\left(
\begin{array}{cc}
E_3&E_{36}\\
E_{36}^{\dag}&E_6
\end{array}
\right)
=\left(
\begin{array}{cc}
D^{\dag}D&D^{\dag}\\
D&I
\end{array}
\right),
\]
where $[D,D^{\dag}]=0$. The matrix $\rho$ has the form:
\[
\rho=\left(
\begin{array}{cccccc}
E_1&E_{12}&E_{13}&E_{14}&E_{15}&E_{16}\\
E_{12}^{\dag}&E_2&E_{23}&E_{24}&E_{25}&E_{26}\\
E_{13}^{\dag}&E_{23}^{\dag}&D^{\dag}D&E_{34}&E_{35}&D^{\dag}\\
E_{14}^{\dag}&E_{24}^{\dag}&E_{34}^{\dag}&C^{\dag}C&C^{\dag}B&C^{\dag}\\
E_{15}^{\dag}&E_{25}^{\dag}&E_{35}^{\dag}&B^{\dag}C&B^{\dag}B&B^{\dag}\\
E_{16}^{\dag}&E_{26}^{\dag}&D&C&B&I
\end{array}
\right).
\]

The above matrix $\rho$ possesses  kernel vectors $|11\ra |f\ra
-|12\ra B|f\ra, \ |10\ra |g\ra -|12\ra C|g\ra$ and $|02\ra |h\ra
-|12\ra D|h\ra$ for all $|f\ra, \ |g\ra$ and $|h\ra$ from the
Charlie's space, which implies that
$E_{34}=D^{\dag}C,\ E_{35}=D^{\dag}B,\ E_{13}=E_{16}D,\
E_{14}=E_{16}C,\ E_{15}=E_{16}B,\ E_{23}=E_{26}D,\
E_{24}=E_{26}C$ and $E_{25}=E_{26}B$. Therefore $\rho$ has the form:
 \[
\rho=\left(
\begin{array}{cccccc}
E_1&E_{12}&E_{16}D&E_{16}C&E_{16}B&E_{16}\\
E_{12}^{\dag}&E_2&E_{26}D&E_{26}C&E_{26}B&E_{26}\\
D^{\dag}E_{16}^{\dag}&D^{\dag}E_{26}^{\dag}&
D^{\dag}D&D^{\dag}C&D^{\dag}B&D^{\dag}\\
C^{\dag}E_{16}^{\dag}&C^{\dag}E_{26}^{\dag}&
C^{\dag}D&C^{\dag}C&C^{\dag}B&C^{\dag}\\
B^{\dag}E_{16}^{\dag}&B^{\dag}E_{26}^{\dag}&
B^{\dag}D&B^{\dag}C&B^{\dag}B&B^{\dag}\\
E_{16}^{\dag}&E_{26}^{\dag}&D&C&B&I
\end{array}
\right).
\]
The partial transpose of $\rho$ with respect to
Alice is given by,
\[
\rho^{t_A}=\left(
\begin{array}{cccccc}
E_1&E_{12}&E_{16}D&C^{\dag}E_{16}^{\dag}&
C^{\dag}E_{26}^{\dag}&C^{\dag}D\\
E_{12}^{\dag}&E_2&E_{26}D&B^{\dag}E_{16}^{\dag}&
B^{\dag}E_{26}^{\dag}&B^{\dag}D\\
D^{\dag}E_{16}^{\dag}&D^{\dag}E_{26}^{\dag}&
D^{\dag}D&E_{16}^{\dag}&E_{26}^{\dag}&D\\
E_{16}C&E_{16}B&E_{16}&C^{\dag}C&C^{\dag}B&C^{\dag}\\
E_{26}C&E_{26}B&E_{26}&B^{\dag}C&B^{\dag}B&B^{\dag}\\
D^{\dag}C&D^{\dag}B&D^{\dag}&C&B&I
\end{array}
\right).
\]
As $\rho^{t_A}$ is positive and any partial transpose
with respect to Alice does
not change $\la1_A|\rho|1_A\ra$, the vectors $|11\ra |f\ra -|12\ra
B|f\ra, \ |10\ra |g\ra -|12\ra C|g\ra$ are still the
kernel vectors. Therefore we have $E_{26}^{\dag}=DB,\
E_{16}^{\dag}=DC$, and $\rho$ becomes
\[
\rho=\left(
\begin{array}{cccccc}
E_1&E_{12}&C^{\dag}D^{\dag}D&C^{\dag}
D^{\dag}C&C^{\dag}D^{\dag}B&C^{\dag}D^{\dag}\\
E_{12}^{\dag}&E_2&B^{\dag}D^{\dag}D&B^{\dag}
D^{\dag}C&B^{\dag}D^{\dag}B&B^{\dag}D^{\dag}\\
D^{\dag}DC&D^{\dag}DB&D^{\dag}D&D^{\dag}C&D^{\dag}B&D^{\dag}\\
C^{\dag}DC&C^{\dag}DB&C^{\dag}D&C^{\dag}C&C^{\dag}B&C^{\dag}\\
B^{\dag}DC&B^{\dag}DB&B^{\dag}D&B^{\dag}C&B^{\dag}B&B^{\dag}\\
DC&DB&D&C&B&I
\end{array}
\right).
\]

Let $X=(E_{12},\ C^{\dag}D^{\dag}D,\ C^{\dag}D^{\dag}C,\
C^{\dag}D^{\dag}B,\ C^{\dag}D^{\dag})$ and
$\rho_5=\Sigma+{\rm diag}(\Delta,\ 0,\ 0,\ 0,\ 0)$,
where
\[
\Sigma=\left(
\begin{array}{ccccc}
B^{\dag}D^{\dag}DB&B^{\dag}D^{\dag}D&B^{\dag}
D^{\dag}C&B^{\dag}D^{\dag}B&B^{\dag}D^{\dag}\\
D^{\dag}DB&D^{\dag}D&D^{\dag}C&D^{\dag}B&D^{\dag}\\
C^{\dag}DB&C^{\dag}D&C^{\dag}C&C^{\dag}B&C^{\dag}\\
B^{\dag}DB&B^{\dag}D&B^{\dag}C&B^{\dag}B&B^{\dag}\\
DB&D&C&B&I
\end{array}
\right),
\]
\be\label{delta}
\Delta=E_2-B^{\dag}D^{\dag}DB,
\ee
diag$(A_1,A_2,...,A_m)$
denotes a diagonal block matrix with blocks $A_1,A_2,...,A_m$.
$\rho$ can then be written in the following
partitioned matrix form:
\[
\rho=\left(
\begin{array}{cc}
E_1&X\\
X^{\dag}&\rho_5
\end{array}
\right).
\]
As $\Sigma$ possesses the following  4N kernel vectors:
$$
\begin{array}{ll}
&(\la f|, \ 0,\ 0,\ 0,-\la f|B^\dag D^\dag )^T,\ \
(0, \la g|,0,\ 0, -\la g|D^\dag )^T,\\[3mm]
&(0,\ 0,\la h|, \ 0,\ -\la h|C^\dag )^T,\ \ (0,\ 0,\ 0,\la i|,\
-\la i|B^\dag )^T
\end{array}
$$
for arbitrary $|f\ra,\ |g\ra,\ |h\ra,\ |i\ra$ in Charlie's
space, so the kernel $K(\Sigma)$ has at least dimension $4N$. On
the other hand $r(\Sigma)+k(\Sigma)=5N$, therefore $r(\Sigma)\leq
N$. While the range of $\Sigma$ has at least dimension $N$ due to
the identity entry on the diagonal. So we have $r(\Sigma)=N$.
Notice that $r(\rho_5)\leq r(\rho)=N$, it is easy to see that
$r(\rho_5)=N$. To show that $\Delta= 0$, we make the following
elementary row transformations on the matrix $\rho_5$,
\be\label{add}
\left(
\begin{array}{ccccc}
I&0&0&0&-B^{\dag}D^{\dag}\\
0&I&0&0&-D^{\dag}\\
0&0&I&0&-C^{\dag}\\
0&0&0&I&-B^{\dag}\\
0&0&0&0&I
\end{array}
\right)\rho_5=
\left(
\begin{array}{ccccc}
\Delta&0&0&0&0\\
0&0&0&0&0\\
0&0&0&0&0\\
0&0&0&0&0\\
DB&D&C&B&I
\end{array}
\right).
\ee
As the rank of $\rho_5$ is $N$, from (\ref{add}) we have
$\Delta=0$, and hence $E_2=B^{\dag}D^{\dag}DB$.

Now, notice that $\la \Psi_f|\rho|\Psi_f\ra=0$ for
$|\Psi_f\ra=|01\ra |f\ra -|12\ra DB|f\ra$ and arbitrary $|f\ra$.
Since $\rho\geq 0$ we have $0=\rho
|\Psi_f\ra=|00\ra |f\ra E_{12}-C^{\dag}D^{\dag}DB |f\ra$, which,
as $|f\ra$ is arbitray, leads to $E_{12}=C^{\dag}D^{\dag}DB$, thus
the matrix $\rho$ becomes
 \[
\rho=\left(
\begin{array}{cccccc}
E_1&C^{\dag}D^{\dag}DB&C^{\dag}D^{\dag}D&C^{\dag}
D^{\dag}C&C^{\dag}D^{\dag}B&C^{\dag}D^{\dag}\\
B^{\dag}D^{\dag}DC&B^{\dag}D^{\dag}DB&B^{\dag}D^{\dag}D&
B^{\dag}D^{\dag}C&B^{\dag}D^{\dag}B&B^{\dag}D^{\dag}\\
D^{\dag}DC&D^{\dag}DB&D^{\dag}D&D^{\dag}C&D^{\dag}B&D^{\dag}\\
C^{\dag}DC&C^{\dag}DB&C^{\dag}D&C^{\dag}C&C^{\dag}B&C^{\dag}\\
B^{\dag}DC&B^{\dag}DB&B^{\dag}D&B^{\dag}C&B^{\dag}B&B^{\dag}\\
DC&DB&D&C&B&I
\end{array}
\right).
\]
The above form can be rewritten as
\[
\ba{rcl}
\rho&=&\left(
\begin{array}{c}
C^{\dag}D^{\dag}\\
B^{\dag}D^{\dag}\\
D^{\dag}\\
C^{\dag}\\
B^{\dag}\\
I
\end{array}
\right)
\left(
\begin{array}{cccccc}
DC&DB&D&C&B&I
\end{array}
\right)
+{\rm diag}(\tilde{\Delta},\ 0,\ 0,\ 0,\ 0,\ 0)\\[15mm]
&\equiv&
\tilde{\Sigma}+{\rm diag}(\tilde{\Delta},\ 0,\ 0,\ 0,\ 0,\ 0),
\ea
\]
where $\tilde{\Delta}=E_1-C^{\dag}D^{\dag}DC$.

$\tilde{\Sigma}$ is PPT and has the following $5N$ kernel vectors:
$$
\begin{array}{ll}
&|00\ra |f\ra-|12\ra DC |f\ra,\ \ |01\ra |g\ra-|12\ra DB |g\ra,
\ \ |02\ra |h\ra-|12\ra D |h\ra,\\[3mm]
&|10\ra |i\ra-|12\ra C |i\ra,\ \ |11\ra |j\ra-|12\ra B |j\ra,
\end{array}
$$
for arbitrary $|f\ra,\ |g\ra,\ |h\ra,\ |i\ra,\ |j\ra$. Similar to
the discussions in the case of $\Delta$ in (\ref{delta}), the matrix
$\tilde{\Delta}$ must vanish and $E_1=C^{\dag}D^{\dag}DC$.
Finally $\rho$ reaches the following form:
\[
\rho=\left(
\begin{array}{c}
C^{\dag}D^{\dag}\\
B^{\dag}D^{\dag}\\
D^{\dag}\\
C^{\dag}\\
B^{\dag}\\
I
\end{array}
\right)
\left(
\begin{array}{cccccc}
DC&DB&D&C&B&I
\end{array}
\right).
\]

The commutative relations $[B,\ D]=[C,\
D]=[B,\ D^{\dag}]=[C,\ D^{\dag}]=0$ follow from the
positivity of all partial transposes of $\rho$.
We first consider
\[
\rho^{t_B}=\left(
\begin{array}{cccccc}
C^{\dag}D^{\dag}DC&B^{\dag}D^{\dag}DC&D^{\dag}DC&
C^{\dag}D^{\dag}C&B^{\dag}D^{\dag}C&D^{\dag}C\\
C^{\dag}D^{\dag}DB&B^{\dag}D^{\dag}DB&D^{\dag}DB&
C^{\dag}D^{\dag}B&B^{\dag}D^{\dag}B&D^{\dag}B\\
C^{\dag}D^{\dag}D&B^{\dag}D^{\dag}D&D^{\dag}D&
C^{\dag}D^{\dag}&B^{\dag}D^{\dag}&D^{\dag}\\
C^{\dag}DC&B^{\dag}DC&DC&C^{\dag}C&B^{\dag}C&C\\
C^{\dag}DB&B^{\dag}DB&DB&C^{\dag}B&B^{\dag}B&B\\
C^{\dag}D&B^{\dag}D&D&C^{\dag}&B^{\dag}&I
\end{array}
\right).
\]
Due to the positivity, the matrix $\rho^{t_B}$ must possess
the kernel vector $|02\ra |h\ra -|12\ra D |h\ra$, which implies
that $[B,\ D]=[C,\ D]=0$. The matrix $\rho^{t_B}$ can be then
written as:
\[
\rho^{t_B}=\left(
\begin{array}{c}
D^{\dag}C\\
D^{\dag}B\\
D^{\dag}\\
C\\
B\\
I
\end{array}
\right)
\left(
\begin{array}{cccccc}
C^{\dag}D&B^{\dag}D&D&C^{\dag}&B^{\dag}&I
\end{array}
\right),
\]
which implies automatically the positivity.

From the positivity of $\rho^{t_{AB}}$,
 \[
\rho^{t_{AB}}=\left(
\begin{array}{cccccc}
C^{\dag}D^{\dag}DC&B^{\dag}D^{\dag}DC&D^{\dag}DC&C^{\dag}DC&B^{\dag}DC&DC\\
C^{\dag}D^{\dag}DB&B^{\dag}D^{\dag}DB&D^{\dag}DB&C^{\dag}DB&B^{\dag}DB&DB\\
C^{\dag}D^{\dag}D&B^{\dag}D^{\dag}D&D^{\dag}D&C^{\dag}D&B^{\dag}D&D\\
C^{\dag}D^{\dag}C&B^{\dag}D^{\dag}C&D^{\dag}DC&C^{\dag}C&B^{\dag}C&C\\
C^{\dag}D^{\dag}B&B^{\dag}D^{\dag}B&D^{\dag}B&C^{\dag}B&B^{\dag}B&B\\
C^{\dag}D^{\dag}&B^{\dag}D^{\dag}&D^{\dag}&C^{\dag}&B^{\dag}&I
\end{array}
\right),
\]
we have that $|02\ra |h\ra
-|12\ra D^{\dag} |h\ra$ is a kernel vector, which results in $[B,\
D^{\dag}]=[C,\ D^{\dag}]=0$. $\rho^{t_{AB}}$ is then of the form:
\[
\rho^{t_{AB}}=\left(
\begin{array}{c}
DC\\
DB\\
D\\
C\\
B\\
I
\end{array}
\right)
\left(
\begin{array}{cccccc}
C^{\dag}D^{\dag}&B^{\dag}D^{\dag}&D^{\dag}&C^{\dag}&B^{\dag}&I
\end{array}
\right).
\]
This form assures positive definiteness, and concludes the
proof of the Lemma. $\Box$

\section{Separability of PPT states supported on $\Cb^2 \otimes
\Cb^3 \otimes \Cb^N$ with rank $N$}

\subsection{Separability of PPT states supported on $\Cb^2 \otimes
\Cb^3 \otimes \Cb^N$ with rank $N\geq 6$}

{\bf Lemma 2.}\ A PPT-state $\rho$ in $\Cb^2 \otimes
\Cb^3 \otimes \Cb^N$ with $r(\rho)=N$, and for which
there exists a product basis $|e_A,\ f_B\ra$, such that $r(\la
e_A,\ f_B|\rho |e_A,\ f_B\ra)=N$, is separable.

{\bf Proof. }\ According to Lemma 1 the state $\rho$ can be written as
\[
\rho=
\left(
\begin{array}{c}
C^{\dag}D^{\dag}\\
B^{\dag}D^{\dag}\\
D^{\dag}\\
C^{\dag}\\
B^{\dag}\\
I
\end{array}
\right)
\left(
\begin{array}{cccccc}
DC&DB&D&C&B&I
\end{array}
\right).
\]
since operators $B$, $C$ and $D$ commute, they have common eigenvectors
$|f_n\ra$, with eigenvalues $b_n,\ c_n$ and $d_n$ respectively. Hence
\[
\la f_n|\rho |f_n\ra=
\left(
\begin{array}{c}
c_n^*d_n^*\\
b_n^*d_n^*\\
d_n^*\\
c_n^*\\
b_n^*\\
1
\end{array}
\right)
\left(
\begin{array}{cccccc}
d_nc_n&d_nb_n&d_n&c_n&b_n&1
\end{array}
\right),
\]
which is a product vector in Alice's and Bob's spaces.
$\rho$ can thus be written as $\rho=\sum_{n=1}^N |\psi_n\ra \la
\psi_n|\otimes |\phi_n\ra \la \phi_n|\otimes |f_n\ra \la f_n|$.
Because the local transformations used above are reversible, we
can apply their inverses and obtain a decomposition of the
initial state $\rho$ in a sum of projecctors onto product vectors.
This proves the separability of $\rho$   $\Box$

We say that $\rho$ acting on $\Cb^2 \otimes\Cb^3 \otimes \Cb^N$
is supported on $\Cb^2 \otimes\Cb^3 \otimes \Cb^N$
if there exist no $M_A<2$, $M_B<3$ and $M_C<N$ such that $R(\rho)\subset
\Cb^{M_A} \otimes\Cb^{M_B} \otimes \Cb^{M_C}$ and $M_A+M_B+M_C<2+3+N$.
Therefore $\rho$ is supported on $\Cb^2 \otimes\Cb^3 \otimes \Cb^N$
if and only if there exists no vector $\vert e\rangle$ in one
of spaces from Alice, Bob, or Charlie, such that $\rho \vert e\rangle=0$.

{\bf Lemma 3.}\ Any PPT state $\rho$ supported on
$\Cb^{2}\otimes \Cb^{3}\otimes \Cb^{N}$ with
$r(\rho)=N$ and $N\ge 6$ is separable, and obeys assumptions of Lemma 1.

{\bf Proof.} A $\Cb^{2}\otimes \Cb^{3}\otimes \Cb^{N}$-system can be
regarded as a $\Cb^{6}\otimes  \Cb^{N}$-system. From the theorem 1 in
\cite{2} we obtain that
\be\label{1}
\rho=|\psi_{AB_1}\ra\la\psi_{AB_1}|\otimes
 |C_1\ra\la C_1|+\sum_{i=2}^N
 |\psi_{AB_i},C_i\ra\la\psi_{AB_i},C_i|.
\ee
Since the vector's $|C_{i}\ra$ are linearly independent, we can
find a vector $|{C}\ra$ in Charlie's space so that $\la C
|\rho|C\ra\sim |\psi_{{AB}_1}\ra\la\psi_{{AB}_1}|$. As the
state $\rho$ has the PPT property with respect to all partitions,
$|\psi_{AB_1}\ra\la\psi_{AB_1}|$ must be PPT with respect to Alice
or Bob (i.e. a product state). This observation
concerns all projectors that enter the convex sum (\ref{1}). Therefore
we conclude that $\rho$ is separable. It follows
directly from (\ref{1}) that $\rho$ can be projected
onto $| 1_A, 2_B\ra$, so that rank $r(\la 1_A,2_B |\rho|1_A,2_B\ra)=N$. $\Box$

For the cases of $N=2,3,4,5$, we consider the separability of states
with different ranks in the following sections.

\subsection{Separability of PPT states supported on $\Cb^2 \otimes
\Cb^2 \otimes \Cb^3$ with rank $N\leq 4$}

{\bf Lemma 4.}\ Any PPT state $\rho$ supported   on
$\Cb^{2}\otimes \Cb^{2}\otimes \Cb^{3}$ with $r(\rho)=2$ is
separable and has a product vector $|e,f,g\ra$ in its kernel.

{\bf Proof.}\ A product vector $|{e,f,g}\ra$ belongs to the kernel iff it
is orthogonal to two vectors $\{|\psi_1\ra,|\psi_2\ra\}$
that span the range  of $\rho$. For arbitrary $|e\ra$ and
$|f\ra=|0\ra+\alpha|1\ra$, we have two equations:
$$
(\la\psi_i|e,0\ra+\alpha(\la\psi_{i}|e,1\ra))|g\ra=0,\quad i=1,2.
$$
We treat these equations as linear homogeneous equations for
$|g\ra$. As the number of equations is smaller than the
number of parameters, the linear homogeneous equations have
always nonzero solution.

Now let $|e_{A},f_{B},g_{C}\ra$ be the kernel vector of $\rho$.
Using PPT property and Lemma 5 in \cite{3}, we have
$\rho^{t_{A}}|{e^{*}_{A},f_{B},g_{C}\ra}=0$. Therefore
$\la{\hat{e}^{*}}_A|\rho^{t_{A}}|{e_A,f_B,g_C}\ra=0$, where and in
the following we denote $|{\hat e}\ra$ to be the vector that is
orthogonal to $|e\ra$. This equation is equivalent to
$\la{e_{A}}|\rho|\hat e_A,f_B,g_C\ra=0$, which implies that
$\rho|\hat e_A,f_B,g_C\ra=|{\hat e_A}\ra|\psi_{BC}\ra$, where
$|\psi_{BC}\ra$ is a vector in Bob's and Charlie's spaces.
According to Lemma 2 in \cite{3} which deals with $\Cb^{2}\otimes
\Cb^{N}$-systems, we can subtract the projector
$|\hat{e}_{A},\psi_{BC}\ra\la\hat{e}_{A},\psi_{BC}|$ from $\rho$,
so that
$$
\tilde{\rho}=\rho- \frac{1}
{\la\hat{e}_A,\psi_{BC}|\rho^{-1}|\hat{e}_A,\psi_{BC}\ra}
|\hat{e}_A,\psi_{BC}\ra\la\hat{e}_A,\psi_{BC}|
$$
is a projector. Since it has the PPT property with
respect to Alice's system, it must be separable with respect to
$A$-$BC$ partition (taking
$A$ to be one of the subsystems of a bipartite system,
and $B$ and $C$ together as another one). In general, we can write
$\rho=\tilde{\Lambda}|\tilde{e}_{A},
\tilde\psi_{BC}\ra\la\tilde{e}_{A},\tilde\psi_{BC}|
+\Lambda|\hat{e}_{A},\psi_{BC}\ra\la\hat{e}_{A},\psi_{BC}|$.
Projecting onto $|e_A\ra$ we get $\la e_A|\rho| e_A\ra\sim
|\tilde\psi_{BC}\ra\la\tilde\psi_{BC}|$. Since $\rho$   has the
PPT property with respect to all partitions, the projectors
$|\tilde\psi_{BC}\ra\la\tilde\psi_{BC}|$ must project onto a
product vector. The same can be said about
$|\psi_{BC}\ra\la\psi_{BC}|$ since the projection onto
$|\hat{\tilde e}_A\ra$ gives $|\hat{\tilde e}_A\ra\la\hat{\tilde
e}_A|\sim|\psi_{BC}\ra\la\psi_{BC}|$, which implies that
$|\psi_{BC}\ra\la\psi_{BC}|$ is a product state and concludes the
proof. $\Box$

{\bf Lemma 5.}\  Any PPT state $\rho$ supported on $\Cb^2\otimes
\Cb^2\otimes \Cb^3$ with $r(\rho)=4$ is separable and has a product vector
$|e,f,g\ra$ in its kernel.

{\bf Proof.}\ A PPT-state $\rho$ in $\Cb^{2}\otimes \Cb^{2}\otimes
\Cb^{3}$ can be regarded as a state in $\Cb^{2}_{A}\otimes
\Cb^{6}_{BC}$. According to Theorem 1 in \cite{3}, this state is supported
on $\Cb^{2}_{A}\otimes \Cb^{4}_{BC}$ and have the form:
$$
\rho=\sum_{i=1}^4 |e_{A_i}\ra\la e_{A_i}|\otimes
|\psi_{BC_i}\ra\la\psi_{BC_i}|.
$$
Let $|{e}\ra$ be orthogonal to $|e_{A_4}\ra$ and
$|{f}\ra=|{0_{B}}\ra+\alpha|{1_{B}}\ra$. By demanding that
$|{f,g}\ra$ is orthogonal to $|{\psi_{BC_{i}}}\ra$ for $i=1,2,3$,
we have the following system of linear homogeneous equations for $|g\ra$:
$$
(\la\psi_{BC_i}|0_B\ra+\alpha\la\psi_{BC_i}|1_B\ra)|g\ra=0\  {
\;\rm for } \;\ i=1,2,3.
$$
These equations possess a nontrivial solution if the corresponding
determinant of the $3\times 3$ matrix vanishes. This leads to a
cubic equation for $\alpha$, which has always a solution. Therefore
$\rho$ has always product kernel vectors.

Let $|{e_{A},f_{B},g_{C}}\ra$ be one of the kernel vectors of $\rho$.
From the condition $\rho|{e_{A},f_{B},g_{C}}\ra=0$, we get,
$$
\la{e_{A}}|\rho|{\hat{e}_{A},f_{B},g_{C}}\ra=0,~~~~
\la{f_{B}}|\rho|{e_{A},\hat{f}_{B},g_{C}}\ra=0,~~~~
\la g_C|\rho|e_A,f_B,\hat{g}_C^i\ra=0,
$$
where
$|\hat{e}_A\ra\perp|e_A\ra$, $|\hat{f}_B\ra\perp|{f}_B\ra$,
$|\hat{g}_C^i\ra\perp|g_C\ra$, $i=1,2$,
$|\hat{g}_C^1\ra\perp|\hat{g}_C^2\ra$. This means that $\rho|{\hat
e_A,f_{B},g_{C}}\ra=|{\hat e_A}\ra|{\psi_{BC}}\ra$,
$\rho|{e_{A},\hat f_B,g_{C}}\ra=|{\hat f_B}\ra|{\psi_{AC}}\ra$,
$\rho|e_A,f_B,\hat g_C^i\ra=|\hat{g}_C^i\ra|\psi_{AB}^i\ra$,
$i=1,2$. We define
\be\label{rho5}
\tilde{\rho}=\rho-\bar\lambda_1|\hat
g_C^1\ra\la\hat g_C^1| \otimes|\psi_{AB}^1\ra\la\psi_{AB}^1|
-\bar\lambda_2|\hat g_C^2\ra\la\hat g_C^2|
\otimes|\psi_{AB}^2\ra\la\psi_{AB}^2|,
\ee
where
$\bar\lambda_i=1/\la\hat
g_C^i,\psi_{AB}^i|\rho^{-1} |\hat g_C^i,\psi_{AB}^i\ra$, $i=1,2$.
$\tilde{\rho}$ is a PPT state with respect to $AB$-$C$
partition (taking $A$ and $B$ together to be one of
the subsystems of a bipartite system,
and $C$ as another one), i.e., $\tilde{\rho}^{t_{C}}\geq 0$,
$r(\tilde{\rho})=2$, from Lemma 2 in \cite{1}. We rewrite
$\tilde{\rho}$ as:
$$
\tilde{\rho}=\lambda_1 |\hat e_A\ra \la\hat e_A | \otimes
|\psi_{BC}\ra\la\psi_{BC}| +\lambda_{2}|\hat f_B\ra\la\hat f_B|
\otimes |\psi_{AC}\ra\la\psi_{AC}|.
$$
Changing the basis in Charlie's system by redefining
$|g_C\ra=|0\ra$, $|\hat g_C^1\ra=|1\ra$, $|\hat g_C^2\ra=|2\ra$,
we have that the vectors $|\psi_{AC}\ra$ and $|\psi_{BC}\ra$ in the
new basis are of the form:
$$
|{\psi_{AC}}\ra=|\psi_A^{1}\ra|0\ra +|\psi_A^{2}\ra|1\ra
+|\psi_A^{3}\ra|2\ra, ~~~~~|{\phi_{BC}}\ra=|\phi_B^{1}\ra|0\ra
+|\phi_B^2\ra|1\ra +|\phi_B^3\ra|2\ra.
$$
Correspondingly $\tilde{\rho}$ can be written as:
$$
\tilde{\rho}=\left(
\begin{array}{ccc}
\lambda_1|\hat e_A\ra\la\hat e_A|\otimes |\phi_B^1\ra\la\phi_B^1|
& \lambda_1|\hat e_A\ra\la\hat e_A|\otimes |\phi_B^1\ra\la\phi_B^2|
& \lambda_1|\hat e_A\ra\la\hat e_A|\otimes
|\phi_B^1\ra\la\phi_B^3|\\
+\lambda_2|\psi_A^1\ra\la\psi_A^1|\otimes |\hat f_B\ra\la\hat f_B|
& +\lambda_2|\psi_A^1\ra\la\psi_A^2|\otimes |\hat f_B\ra\la\hat f_B|
& +\lambda_2|\psi_A^1\ra\la\psi_A^3|\otimes |\hat f_B\ra\la\hat
f_B|\\[3mm]
\lambda_1|\hat e_A\ra\la\hat e_A|\otimes |\phi_B^2\ra\la\phi_B^1|
&\lambda_1|\hat e_A\ra\la\hat e_A|\otimes |\phi_B^2\ra\la\phi_B^2|
& \lambda_1|\hat e_A\ra\la\hat e_A|\otimes
|\phi_B^2\ra\la\phi_B^3|\\
+\lambda_2|\psi_A^2\ra\la\psi_A^1|\otimes |\hat f_B\ra\la\hat f_B|
&+\lambda_2|\psi_A^2\ra\la\psi_A^2|\otimes |\hat f_B\ra\la\hat f_B|
&+\lambda_2|\psi_A^2\ra\la\psi_A^3|\otimes |\hat f_B\ra\la\hat
f_B|\\[3mm]
\lambda_1|\hat e_A\ra\la\hat e_A|\otimes |\phi_B^3\ra\la\phi_B^1|
& \lambda_1|\hat e_A\ra\la\hat e_A|\otimes |\phi_B^3\ra\la\phi_B^2|
& \lambda_1|\hat e_A\ra\la\hat e_A|\otimes
|\phi_B^3\ra\la\phi_B^3|\\
+\lambda_2|\psi_A^3\ra\la\psi_A^1|\otimes |\hat f_B\ra\la\hat f_B|
& +\lambda_2|\psi_A^3\ra\la\psi_A^2|\otimes |\hat f_B\ra\la\hat
f_B| & +\lambda_2|\psi_A^3\ra\la\psi_A^3|\otimes |\hat
f_B\ra\la\hat f_B|
\end{array}
\right).$$

From the positivity of $\tilde{\rho}$ and $\tilde{\rho}^{t_{c}}$,
if the last column acting on
$|\hat{\psi}_A^3\ra|\hat{\phi}_B^3\ra$ vanishes, the same must be
true for the last row. Similarly, if the second (resp. first) column acting
on $|\hat{\psi}_A^2\ra|\hat{\phi}_B^2\ra$
(resp. $|\hat{\psi}_A^1\ra|\hat{\phi}_B^1\ra$) vanishes, the same must
be true for the corresponding rows. This leads to a set of equations:
$$
\la\hat e_A|\hat{\psi}_A^i\ra\la\phi_B^j|\hat\phi_B^i\ra=0,~~~~
\la\psi_A^j|\hat{\psi}_A^i\ra\la\hat f_B|\hat\phi_B^i\ra=0,~~~~
i=1,2,3,~~~ j\not=i.
$$
This equation set implies that at least one of the
projectors of $|\psi_{BC}\ra\la\psi_{BC}|$ and
$|\psi_{AC}\ra\la\psi_{AC}|$ must be a product state. If it is,
for instance, $|\psi_{AC}\ra\la\psi_{AC}|$, then
$|\psi_A^1\ra=|\psi_A^2\ra =|\psi_A^3\ra =|\hat e_A\ra$,
$|\psi_{AC}\ra=|\hat e_A\ra|\tilde{g}_C\ra$, where
$|\tilde{g}_C\ra=|0\ra+|1\ra+|2\ra$ and  $\rho$ becomes
$$
\ba{rcl}
\rho&=&\lambda_1|\hat e_A\ra\la\hat e_A|\otimes
|\psi_{BC}\ra\la\psi_{BC}| +\lambda_2|\hat e_A\ra\la\hat e_A|\otimes
|\hat f_B\ra\la\hat f_B|\otimes
|\tilde{g}_C\ra\la\tilde{g}_C|\\[3mm]
&&+\bar{\lambda}_1|\hat{g}_C^1\ra\la\hat{g}_C^1|\otimes
|\psi_{AB}^1\ra\la\psi_{AB}^1|
+\bar{\lambda}_2|\hat{g}_C^2\ra\la\hat{g}_C^2|\otimes
|\psi_{AB}^2\ra\la\psi_{AB}^2|.
\ea
$$
Let $\sigma=\lambda_1 |\psi_{BC}\ra\la\psi_{BC}| +\lambda_2 |\hat
f_B\ra\la\hat f_B|\otimes|\tilde g_C\ra\la\tilde g_C|$. The
operator $\sigma$ is a PPT state of rank $2$ in $\Cb^2 \otimes
\Cb^3$  spaces of Bob and Charlie. From Peres-Horodecki criterion
it is separable. The matrix $\rho$ can thus be written as
$$
\ba{rcl} \rho&=&\lambda_1|\hat{e}_A\ra\la\hat{e}_A|\otimes
|\tilde{f}_B\ra\la\tilde{f}_B|\otimes |\bar{g}_C\ra\la\bar{g}_C|
+\lambda_2|\hat{e}_A\ra\la\hat{e}_A|\otimes
|\hat{f}_B\ra\la\hat{f}_B|\otimes
 |\tilde{g}_C\ra\la\tilde{g}_C|\\[3mm]
&&+\bar{\lambda}_1|\hat{g}_C^1\ra\la\hat{g}_C^1|\otimes
 |\psi_{AB}^1\ra\la\psi_{AB}^1|
 +\bar{\lambda}_2|\hat{g}_C^2\ra\la\hat{g}_C^2|\otimes
 |\psi_{AB}^2\ra\la\psi_{AB}^2|.
\ea
$$

We can also write
$$
\check{\rho}=\rho-
\bar{\lambda}|\hat{e}_A\ra\la\hat{e}_A|\otimes
|\tilde{f}_B\ra\la\tilde{f}_B|\otimes |\bar{g}_C\ra\la\bar{g}_C|
-\check{\lambda}|\hat{e}_A\ra\la\hat{e}_A|\otimes
|\hat{f}_B\ra\la\hat{f}_B|\otimes
 |\tilde{g}_C\ra\la\tilde{g}_C|,
$$
That is
$$
\check{\rho}=
\bar{\lambda}_1|\hat{g}_C^1\ra\la\hat{g}_C^1|\otimes
|\psi_{AB}^1\ra\la\psi_{AB}^1|
+\bar{\lambda}_2|\hat{g}_C^2\ra\la\hat{g}_C^2|\otimes
|\psi_{AB}^2\ra\la\psi_{AB}^2|.
$$
The projection of $\check{\rho}$ onto $|\hat{g}_C^1\ra$ gives
$\la\hat{g}_C^1|\check{\rho}|\hat{g}_C^1\ra \sim
|\psi_{AB}^1\ra\la\psi_{AB}^1|$. This  means that
$|\psi_{AB}^1\ra\la\psi_{AB}^1|$ is a PPT state and hence
$|\psi_{AB}^1\ra$ must be a product vector. Similar discussions
apply also to the state $|\psi_{AB}^2\ra$.  $\Box$

\subsection{Separability of PPT states supported on $\cb^2 \otimes
\cb^3 \otimes \cb^3$ with rank$=3,4$}

{\bf Lemma 6.}\ Any PPT state $\rho$ supported on
$\Cb^{2}\otimes \Cb^{3}\otimes \Cb^{3}$ with $r(\rho)=3$ is
separable.

{\bf Proof.}\ We consider the system $\Cb^{2}\otimes \Cb^{3}\otimes
\Cb^{3}$-system as a $\Cb_{AB}^6 \otimes \Cb_C^3$-system. According to
the Theorem 1 in \cite{2}, threre are possibilities:
(i). $\rho$ is supported on $\Cb_{AB}^3\otimes \Cb_C^3$ and of
the form:
$$
\rho=\sum_{i=1}^3\lambda_i|e_{AB_i}\ra\la e_{AB_i}|\otimes
|f_{C_i}\ra\la f_{C_i}|.
$$
Since the vectors $|f_{C_i}\ra$ are
linearly independent, we can find a vector $|C^i\ra$ in Charlie's
system, such that $\la {C^i}|\rho|{C^i}\ra\sim |e_{AB_i}\ra\la
e_{AB_i}|,\  i=1,2,3$. Because the considered  state $\rho$ has
the PPT property with respect to all partitions, the projected
$|e_{AB_i}\ra\la e_{AB_i}|$ is also PPT, and as such must be a
product state.
(ii). $\rho$ is supported on $\Cb_{AB}^2\otimes \Cb_C^3$. The same
method of projecting onto appropriately chosen vector in third
space as in $(i)$ applies.
(iii). $\rho$ is supported on $\Cb_{AB}^3\otimes
\Cb_C^2$. That is nothing else but a state in $\Cb^2\otimes
\Cb^2\otimes \Cb^3$-system with rank $3$. Its separability follows
from Lemma $6$ in \cite{2}. $\Box$

{\bf Lemma 7.}\ Any PPT state $\rho$ supported on $
\Cb^{2}\otimes \Cb^{3}\otimes \Cb^{3}$ with $r(\rho)=4$ is separable
and has a product vector $|{e,f,g}\ra$ in its kernel.

{\bf Proof.}\ A PPT-state $\rho$ in $\Cb^{2}\otimes \Cb^{3}\otimes
\Cb^{3}$ can be regarded as a state in $\Cb^{6}_{AB}\otimes
\Cb^{3}_{C}$. According to Theorem 1 in \cite{2}, this state is supported
on $\Cb^{4}_{AB}\otimes  \Cb^{3}_{C}$ and have the form:
$$
\rho=\sum_{i=1}^4 |\psi_{AB_i}\ra\la \psi_{AB_i}| \otimes
|g_{C_i}\ra\la g_{C_i}|.
$$

We take $|g\ra$ orthogonal to $|g_{C_4}\ra$, and demand that
$|e,f\ra$ is orthogonal to $|\phi_{AB_i}\ra$, $i=1,2,3$. Setting
$|e_A\ra=|0_A\ra+\alpha|1_A\ra$, we obtain the following system of
linear homogeneous equations for $|f\ra$:
$$
\la\phi_{AB_i}|0_A\ra+\alpha\la\phi_{AB_i}|1_A\ra)|f\ra=0,~~~~
\forall i=1,~2,~3.
$$
These equations possesses a nontrivial solution if the corresponding
determinant of the $3\times 3$ matrix vanishes. The
cubic equation for $\alpha$ has always a solution
and hence a product kernel vector
of $\rho$ exists.

From the kernel product vector,
similar to Lemma 5, one can prove that $\rho$ is
separable, by noting that
$\rho$ is a PPT state of rank $2$ in $\Cb^3\otimes \Cb^3$ and using the
result of \cite{3}.
We also can substract two terms from the second system as in
Lemma 5 from the third one. $\Box$

\subsection{Separability of PPT states with rank$=4$
in $\Cb^2 \otimes \Cb^3 \otimes \Cb^4$}

{\bf Lemma 8.}\ Any PPT state $\rho$ supported on
$\Cb^{2}\otimes \Cb^{3}\otimes \Cb^{4}$ with $r(\rho)=4$ is
separable.

{\bf Proof.} We consider the system $\Cb^{2}\otimes \Cb^{3}\otimes
\Cb^{4}$-system as a $\Cb_{AB}^6 \otimes \Cb_C^4$-system. According to
the Theorem 1 in \cite{1} there are five possibilities:

$1)$\ The state is supported on $\Cb_{AB}^4\otimes \Cb_C^4$. In this case the
density matrix must have the form:
$$
\rho=\sum_{i=1}^4\lambda_i|e_{AB_i}\ra\la e_{AB_i}|\otimes
|f_{C_i}\ra\la f_{C_i}|.
$$
Since the vectors $|f_{C_i}\ra$ are linearly independent, we can
find a vector $|C^i\ra$ in Charlie's system, such that $\la
{C^i}|\rho|{C^i}\ra\sim |e_{AB_i}\ra\la e_{AB_i}|$, $i=1,2,3,4$.
Because $\rho$ has the PPT property with
respect to all partitions, the projected $|e_{AB_i}\ra\la
e_{AB_i}|$ is also PPT , and as such must be a product state.

$2)$\ The state is supported on $\Cb_{AB}^3\otimes \Cb_C^4$. The same
method used in $1)$ applies.

$3)$\  The state is supported on $\Cb_{AB}^2\otimes \Cb_C^4$. The same
method  used in $1)$ applies.

$4)$\  The state is supported on $\Cb_{AB}^4\otimes \Cb_C^3$. That is
just a state in $\Cb^2\otimes \Cb^3\otimes
\Cb^3$-system with rank $4$. Its separability follows from
Lemma 7.

$5)$ \  The state is supported on $\Cb_{AB}^4\otimes \Cb_C^2$. It is a
state in $\Cb^2\otimes \Cb^2\otimes \Cb^3$-system with rank $4$. Its
separability follows from Lemma 5. $\Box$

\subsection{Separability of PPT states with rank$=5$
in $\Cb^2 \otimes \Cb^3 \otimes \Cb^5$}

{\bf Lemma 9.}\ Any PPT state $\rho$ supported on
$\Cb^{2}\otimes \Cb^{3}\otimes \Cb^{5}$ with $r(\rho)=5$ is separable and
has a product vector $|{e,f,g}\ra$ in the kernel.

{\bf Proof.}\ The vector $|{e,f,g}\ra$ belongs to the kernel iff it
is orthogonal to five vectors $\{|\psi_i\ra\}$, $i=1,2,3,4,5$, that span the range
of $\rho$. By choosing arbitrary $|f\ra$ and
$|e\ra=|0\ra+\alpha|1\ra$, we obtain five equations:
$$
(\la\psi_i|0,f\ra+\alpha\la\psi_i|1,f\ra)|g\ra=0,\ \  i=1,2,3,4,5.
$$
We treat these equations as linear homogeneous equations for
$|g\ra$. These equations possess a nontrivial solution if the
corresponding determinant of the $5\times 5$ matrix vanishes. This
leads to a quintic equation for $\alpha$, which has always a
solution and such product kernel vectors exist.

Let $|{e_{A},f_{B},g_{C}}\ra$ be one of the
kernel vectors of $\rho$. Using the condition $\rho|{e,f,g}\ra=0$,
we have
$$
\ba{l}
\rho|{\hat e_A,f_{B},g_{C}}\ra=|{\hat e_A}\ra|{\psi_{BC}}\ra,~~~~~
\rho|{e_{A},\hat f_B,g_{C}}\ra=|{\hat f_B}\ra|{\psi_{AC}}\ra,\\[4mm]
\rho|e_A,f_B,\hat g_C^i\ra=|\hat g_C^i\ra|\psi_{AB}^i\ra,~~~~i=1,~2,~3,~4,
\ea
$$
where $|\hat e_A\ra\perp|e_A\ra$, $\hat
f_B\ra\perp|f_B\ra$, $|\hat g_C^i\ra\perp|g_C\ra$, $|\hat
g_C^k\ra\perp|\hat g_C^l\ra$, $i,k=1,2,3$, $k\not=l$.

We define
$\tilde{\rho}=\rho-\sum_{i=1}^3\bar{\lambda}_i|\hat{g}_C^i\ra\la\hat{g}_C^i|
\otimes |\psi_{AB}^i\ra\la\psi_{AB}^i|$, where
$\bar{\lambda}_i=(\la\hat g_C^i,\psi_{AB}^i|\rho^{-1}
|\hat g_C^i, \psi_{AB}^i\ra)^{-1}$, $i=1,2,3$. $\tilde{\rho}$ is a
PPT state with respect to $AB$-$C$ partition, i.e.,
$\tilde{\rho}^{t_C}\geq 0$, $r(\tilde{\rho})=2$. Using Lemma 2
in \cite{1}, we get
$$
\tilde{\rho}=\lambda_{1}|\hat e_A\ra\la\hat e_A|\otimes
|\psi_{BC}\ra\la\psi_{BC}| +\lambda_{2}|\hat f_B\ra\la\hat
f_B|\otimes |\psi_{AC}\ra\la\psi_{AC}|.
$$
By changing the basis in Charlie's system,
$|g_C\ra=|0\ra$, $|\hat g_C^1\ra=|1\ra$, $|\hat
g_C^2\ra=|2\ra$, $|\hat g_C^3\ra=|3\ra$, $|\hat g_C^4\ra=|4\ra$,
we get new representations of the vectors
$|\psi_{AC}\ra$ and $|\psi_{BC}\ra$:
$$
\ba{l}
|\psi_{AC}\ra=|\psi_A^1\ra|0\ra +|\psi_A^2\ra|1\ra
+|\psi_A^3\ra|2\ra +|\psi_A^4\ra|3\ra +|\psi_A^5\ra|4\ra,\\[4mm]
|\psi_{BC}\ra=|\phi_B^1\ra|0\ra +|\phi_B^2\ra|1\ra
+|\phi_B^3\ra|2\ra +|\phi_B^4\ra|3\ra +|\phi_B^5\ra|4\ra.
\ea
$$

In the matrix form $\tilde{\rho}$ can be written as:
$\tilde{\rho}=\left(\begin{array}{cc}
\rho_1&X\\
Y&Z\end{array}\right)$,
where $\rho_1 =\tilde{\rho}$ given in (\ref{rho5}),
$$
X=\left(
\begin{array}{ccc}
\lambda_1|\hat e_A\ra\la\hat e_A|\otimes |\phi_B^1\ra\la\phi_B^4|
& \lambda_1|\hat e_A\ra\la\hat e_A|\otimes |\phi_B^2\ra\la\phi_B^4|
& \lambda_1|\hat e_A\ra\la\hat e_A|\otimes
|\phi_B^3\ra\la\phi_B^4|\\
+\lambda_2|\psi_A^1\ra\la\psi_A^4|\otimes |\hat f_B\ra\la\hat f_B|
& +\lambda_2|\psi_A^2\ra\la\psi_A^4|\otimes |\hat f_B\ra\la\hat f_B|
& +\lambda_2|\psi_A^3\ra\la\psi_A^4|\otimes |\hat f_B\ra\la\hat
f_B|\\[3mm]
\lambda_1|\hat e_A\ra\la\hat e_A|\otimes |\phi_B^1\ra\la\phi_B^5|
&\lambda_1|\hat e_A\ra\la\hat e_A|\otimes |\phi_B^2\ra\la\phi_B^5|
& \lambda_1|\hat e_A\ra\la\hat e_A|\otimes
|\phi_B^3\ra\la\phi_B^5|\\
+\lambda_2|\psi_A^1\ra\la\psi_A^5|\otimes |\hat f_B\ra\la\hat f_B|
&+\lambda_2|\psi_A^2\ra\la\psi_A^5|\otimes |\hat f_B\ra\la\hat f_B|
&+\lambda_2|\psi_A^3\ra\la\psi_A^5|\otimes |\hat f_B\ra\la\hat
f_B|
\end{array}
\right),
$$
$$
Y=\left(
\begin{array}{ccc}
\lambda_1|\hat e_A\ra\la\hat e_A|\otimes |\phi_B^4\ra\la\phi_B^1|
& \lambda_1|\hat e_A\ra\la\hat e_A|\otimes |\phi_B^4\ra\la\phi_B^2|
& \lambda_1|\hat e_A\ra\la\hat e_A|\otimes
|\phi_B^4\ra\la\phi_B^3|\\
+\lambda_2|\psi_A^4\ra\la\psi_A^1|\otimes |\hat f_B\ra\la\hat f_B|
& +\lambda_2|\psi_A^4\ra\la\psi_A^2|\otimes |\hat f_B\ra\la\hat f_B|
& +\lambda_2|\psi_A^4\ra\la\psi_A^3|\otimes |\hat f_B\ra\la\hat
f_B|\\[3mm]
\lambda_1|\hat e_A\ra\la\hat e_A|\otimes |\phi_B^5\ra\la\phi_B^1|
&\lambda_1|\hat e_A\ra\la\hat e_A|\otimes |\phi_B^5\ra\la\phi_B^2|
& \lambda_1|\hat e_A\ra\la\hat e_A|\otimes
|\phi_B^5\ra\la\phi_B^3|\\
+\lambda_2|\psi_A^5\ra\la\psi_A^1|\otimes |\hat f_B\ra\la\hat f_B|
&+\lambda_2|\psi_A^5\ra\la\psi_A^2|\otimes |\hat f_B\ra\la\hat f_B|
&+\lambda_2|\psi_A^5\ra\la\psi_A^3|\otimes |\hat f_B\ra\la\hat
f_B|
\end{array}
\right),
$$
$$
Z=\left(
\begin{array}{cc}
\lambda_1|\hat e_A\ra\la\hat e_A|\otimes |\phi_B^4\ra\la\phi_B^4|
& \lambda_1|\hat e_A\ra\la\hat e_A|\otimes |\phi_B^4\ra\la\phi_B^5|\\
+\lambda_2|\psi_A^4\ra\la\psi_A^4|\otimes |\hat f_B\ra\la\hat f_B|
& +\lambda_2|\psi_A^4\ra\la\psi_A^5|\otimes |\hat f_B\ra\la\hat f_B|
\\[3mm]
\lambda_1|\hat e_A\ra\la\hat e_A|\otimes |\phi_B^5\ra\la\phi_B^4|
&\lambda_1|\hat e_A\ra\la\hat e_A|\otimes |\phi_B^5\ra\la\phi_B^5|\\
+\lambda_2|\psi_A^5\ra\la\psi_A^4|\otimes |\hat f_B\ra\la\hat f_B|
&+\lambda_2|\psi_A^5\ra\la\psi_A^5|\otimes |\hat f_B\ra\la\hat
f_B|
\end{array}
\right).$$

From the positivity of $\tilde{\rho}$ and $\tilde{\rho}^{t_{c}}$
we have that when the $i$-th column acting on
$|\hat{\psi_{A}^{i}}\ra|\hat{\phi_{B}^{i}}\ra$ vanishes, the same
must be true for the corresponding row. This leads to the
equation set:
$$
\la\hat e_A|\hat{\psi}_A^i\ra\la\phi_B^j|\hat\phi_B^i\ra=0,\quad
\la\psi_A^j|\hat{\psi}_A^i\ra\la\hat f_B|\hat\phi_B^i\ra=0,\
~~~i=1,2,3,4,5,\ ~~~j\not=i.
$$
This equation set implies that at least one of the
projectors, $|\psi_{BC}\ra\la\psi_{BC}|$ or
$|\psi_{AC}\ra\la\psi_{AC}|$, must be a product state. If it is,
for instance, $|\psi_{BC}\ra\la\psi_{BC}|$, then
$|\phi_B^1\ra=|\phi_B^2\ra =|\phi_B^3\ra =|\phi_B^4\ra =|\hat
f_B\ra$, $|\psi_{BC}\ra=|\hat f_B\ra|\tilde{g}_C\ra$, where
$|\tilde{g}_C\ra=|0\ra+|1\ra+|2\ra+|3\ra+|4\ra$ and  $\rho$
becomes
$$
\ba{rcl}
\rho&=&\lambda_1|\hat{e}_A\ra\la\hat{e}_A|\otimes
|\hat{f}_B\ra\la\hat{f}_B|\otimes |\tilde{g}_C\ra\la\tilde{g}_C|
+\lambda_2|\hat f_B\ra\la\hat
f_B|\otimes|\psi_{AC}\ra\la\psi_{AC}|\\[4mm]
&&+\sum_{i=1}^3\bar{\lambda_i}|\hat{g}_C^i\ra\la\hat{g}_C^i|\otimes
|\psi_{AB}^i\ra\la\psi_{AB}^i|\\[4mm]
&=&
|\hat{f}_B\ra\la\hat{f}_B|\otimes 
\sigma+\sum_{i=1}^3\bar{\lambda_i}|\hat{g}_C^i\ra\la\hat{g}_C^i|\otimes
|\psi_{AB}^i\ra\la\psi_{AB}^i|,
\ea
$$
where $\sigma=\lambda_1 |\hat e_A\ra\la\hat e_A|\otimes|\tilde
g_C\ra\la\tilde g_C| +\lambda_2 |\psi_{AC}\ra\la\psi_{AC}|$. The
operator $\sigma$ is a PPT state of rank $2$ in $\Cb^2 \otimes
\Cb^5$. Therefore $\sigma$ is separable and can be written in a form
$$
\sigma=\lambda_1 |\hat{e}_A\ra\la\hat{e}_A|\otimes |\tilde
g_C\ra\la\tilde
g_C|+\lambda_2|\tilde{e}_A\ra\la\tilde{e}_A|\otimes
|\bar{g}_C\ra\la\bar{g}_C|.
$$
The matrix $\rho$ can thus be written as
$$
\rho=\lambda_1|\hat e_A\ra\la\hat e_A|\otimes
|\hat{f}_B\ra\la\hat{f}_B|\otimes |\tilde{g}_C\ra\la\tilde{g}_C
+\lambda_2|\tilde e_A\ra\la\tilde e_A|\otimes |\hat f_B\ra\la\hat
f_B|\otimes |\bar g_C\ra\la\bar g_C|
+\sum_{i=1}^3\bar{\lambda}_i|\hat{g}_C^i\ra\la\hat{g}_C^i|\otimes
|\psi_{AB}^i\ra\la\psi_{AB}^i|.
$$

We can also write
$$
\check{\rho}=\rho- \bar\lambda|\hat{e_A}\ra\la\hat{e_A}|\otimes
|\hat{f}_B\ra\la\hat{f}_B|\otimes |\tilde{g}_C\ra\la\tilde{g}_C|
-\check\lambda|\tilde{e}_A\ra\la\tilde{e}_A|\otimes
|\hat{f}_B\ra\la\hat{f}_B|\otimes |\bar{g}_C\ra\la\bar{g}_C|,
$$
where $\bar{\lambda}\equiv \lambda_1
=(\la\hat{e}_A,\hat{f}_B,
\tilde{g}_C|\rho^{-1}|\hat{e}_A,\hat{f}_B, \tilde{g}_C\ra)^{-1}$,
$\check{\lambda}\equiv \lambda_2
=(\la\tilde{e}_A,\hat{f}_B,
\bar{g}_C|\rho^{-1}|\tilde{e}_A,\hat{f}_B, \bar{g}_C\ra)^{-1}$ and
$\check{\rho}$  is a PPT state with respect to $B$-$AC$ partition,
$\check{\rho}=\sum_{i=1}^3 \bar{\lambda}_i|\hat{g}_C^i\ra\la\hat{g}_C^i|\otimes
|\psi_{AB}^i\ra\la\psi_{AB}^i|$.
The projection of $\check{\rho}$ onto $|\hat{g}_C^i\ra$ gives rise to
$\la\hat{g}_C^i|\check{\rho}|\hat{g}_C^i\ra \sim
|\psi_{AB}^i\ra\la\psi_{AB}^i|$, $i=1,~2,~3$. This implies
that $|\psi_{AB}^i\ra$ must be a product vector. $\Box$

By summarizing the above results we have the following conclusion:

{\bf Theorem.}\  Every PPT state $\rho$, supported on
$\Cb^{2}\otimes \Cb^{3}\otimes \Cb^{N}$ with $r(\rho)=N$
is separable, and has a canonical form (\ref{lemma1}).

\section{Remarks}

We have studied the separability and entanglement of quantum
mixed states in  $\Cb^2 \otimes \Cb^3 \otimes \Cb^N$ composite quantum systems.
It is shown that all quantum states $\rho$
supported on $\Cb^{2}\otimes \Cb^{3}\otimes \Cb^{N}$ with positive partial transposes
and rank $r(\rho)\leq N$ are separable. Comparing with the case of
$\Cb^2 \otimes \Cb^2 \otimes \Cb^N$,
the separability and entanglement of mixed states
in $\Cb^2 \otimes \Cb^3 \otimes \Cb^N$ is more complicated. But quite similar
results exist in both cases. Nevertheless, we find that it could be rather
difficult to generalize the results to higher dimensional case, e.g., for states
in $\Cb^3 \otimes \Cb^3 \otimes \Cb^N$ or $\Cb^2 \otimes \Cb^4 \otimes \Cb^N$, as
the PPT criterion is only necessary and sufficient for separability
of states in $\Cb^2 \otimes \Cb^2$ and $\Cb^2 \otimes \Cb^3$
composite quantum systems.

\vskip1cm
\noindent{\bf Acknowledgement} This work is supported by NSF of
China (No. 19975061) and the National Key Project for Basic
Research of China (G1998030601).

\end{document}